\documentclass[twocolumn,showpacs]{revtex4-1} 

\pdfoutput=1

\usepackage{amsmath,amssymb}
\usepackage{mathtools}
\usepackage{graphicx}
\usepackage{bm}

\usepackage{color}

\newcommand{\bea}{\begin{eqnarray}}
\newcommand{\eea}{\end{eqnarray}}

\newcommand{\tm}{\tilde{m}}
\newcommand{\ts}{\tilde{\sigma}}

\newcommand{\dmi}{{\delta m_i}}

\newcommand{\ta}{\tilde{a}}
\newcommand{\teps}{\tilde{\varepsilon}}

\begin{document}

\title{Dynamics of a linear magnetic ``microswimmer molecule''}

\author{Sonja Babel}
\affiliation{                    
  Institut f\"ur Theoretische Physik II: Weiche Materie,
  Heinrich-Heine-Universit\"at D\"{u}sseldorf,
  Universit{\"a}tsstra{\ss}e 1, D-40225 D\"{u}sseldorf, Germany
}

\author{Hartmut L\"owen}
\affiliation{                    
  Institut f\"ur Theoretische Physik II: Weiche Materie,
  Heinrich-Heine-Universit\"at D\"{u}sseldorf,
  Universit{\"a}tsstra{\ss}e 1, D-40225 D\"{u}sseldorf, Germany
}

\author{Andreas M. Menzel}
\affiliation{                    
  Institut f\"ur Theoretische Physik II: Weiche Materie,
  Heinrich-Heine-Universit\"at D\"{u}sseldorf,
  Universit{\"a}tsstra{\ss}e 1, D-40225 D\"{u}sseldorf, Germany
}


\begin{abstract}
In analogy to nanoscopic molecules that are composed of individual atoms, we consider an active ``microswimmer molecule''. It is built up from three individual magnetic colloidal microswimmers that are connected by harmonic springs and hydrodynamically interact with each other. In the ground state, they form a linear straight molecule. We analyze the relaxation dynamics for perturbations of this straight configuration. As a central result, with increasing self-propulsion, we observe an oscillatory instability in accord with a subcritical Hopf bifurcation scenario. It is accompanied by a corkscrew-like swimming trajectory of increasing radius. Our results can be tested experimentally, using for instance magnetic self-propelled Janus particles, supposably linked by DNA molecules. 
\end{abstract}

\maketitle

%
%
%
%

\section{Introduction}
\label{sec:intro}

Often, self-propelled objects are realized on the colloidal level in the form of active microswimmers \cite{menzel2015tuned,elgeti2015physics}. Examples are Janus particles selectively heated \cite{jiang2010active,buttinoni2012active} or catalyzing chemical reactions \cite{howse2007self,theurkauff2012dynamic} on one of their hemispheres, or representatives of nature in the form of swimming microorganisms \cite{berg1975bacterial}. 
As witnessed by several reviews \cite{cates2012diffusive,romanczuk2012active,menzel2015tuned,elgeti2015physics}, the migration behavior of individual self-propelled particles has been studied intensely. If not guided from outside, the long-term translation dynamics of individual self-propelled particles appears diffusive due to fluctuations \cite{howse2007self}. 
In contrast to that, interactions between many self-propelled objects can induce directed collective motion \cite{toner1995long,vicsek1995novel,gregoire2004onset}. 
Steric \cite{wensink2012meso,weber2013long} or elastic \cite{menzel2013traveling,ferrante2013collective,ferrante2013elasticity,menzel2014active} interactions are sufficient for this purpose. 

It is now time to extend the hitherto conception, where individual microswimmers serve as the immediate building blocks of active matter, to a more hierarchical approach. 
In passive equilibrium, just as atoms form nanoscopic molecules, individual colloidal particles were combined to ``colloidal molecules'' \cite{manoharan2003dense,blaaderen2003chemistry,kraft2009colloidal}. Here, we address active systems. We introduce the effect of \textit{permanent elastic bonds} between individual microswimmers, such that they form a self-propelled colloidal ``\textit{microswimmer molecule}''. This is different from 
phoretically stabilized aggregates of catalytically active, not necessarily self-propelled colloidal particles 
\cite{soto2014self,soto2015self,liebchen2015clustering} or clusters of active dipolar particles \cite{kaiser2015active}. Even when our individual swimmers try to drive themselves apart, the bonds hold them together. In the end, the macroscopic behavior of systems composed of many such microswimmer molecules may be analyzed. 

At present, we investigate the stability of the directed motion of one single colloidal microswimmer molecule. Three individual microswimmers are linearly connected 
by harmonic springs, see Fig.~\ref{fig:sketch}. The swimmers form a straight arrangement in the unperturbed equilibrium ground state, stabilized by dipolar magnetic interactions.
Individual swimmers propel by setting the surrounding fluid into motion. They act on their environment in the form of force dipoles \cite{yeomans2014introduction}, i.e.\ two little-distanced force centers that apply antiparallel forces of equal magnitude onto the surrounding fluid, setting it into motion. For non-symmetric swimmer geometries, the self-induced fluid flows can drag the swimmers forward \cite{baskaran2009statistical,menzel2015dynamical}. 
Through the self-induced fluid flows, the three swimmers interact with each other hydrodynamically. 
As a central result, we find that straight configurations of the swimmer molecule and straight trajectories become unstable above a certain strength of self-propulsion. Above this threshold, an oscillatory (Hopf-like) instability resulting in corkscrew-like motion arises.

\section{Model}\label{sec:model}

\begin{figure}
\begin{center}
\includegraphics[width=0.8\columnwidth]{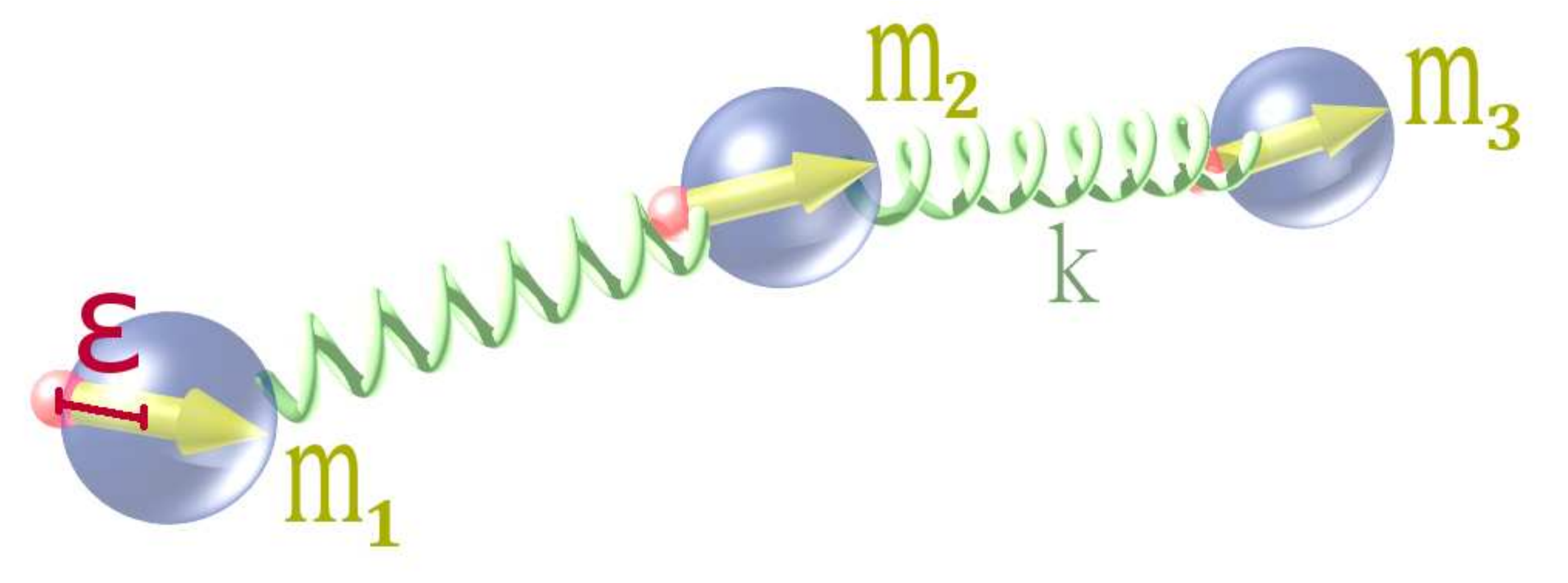}
\caption{
Simplified magnetic colloidal microswimmer molecule, 
here slightly perturbed away from the aligned straight ground state. The three active swimmers are linearly linked by harmonic springs of spring constant $k$. On each swimmer, an active force dipole acts on the surrounding fluid, shifted out of the center by a distance $\varepsilon$ along the vector $\mathbf{m}_i$ ($i=1,2,3$). The latter marks the magnetic dipole moment. In the depicted case of $\varepsilon<0$ we set $\sigma_0>0$ for the strength of the active drive (pusher); for $\varepsilon>0$ the active centers are shifted towards the heads of the vectors $\mathbf{m}_i$ and we set $\sigma_0<0$ (puller).  
\label{fig:sketch}}
\end{center}
\end{figure}

Three identical spherical colloidal microswimmers are labeled by $i=1,2,3$, see Fig.~\ref{fig:sketch}. Each carries a permanent magnetic dipole moment $\mathbf{{m}}_i$ of equal and constant magnitude $m$. The swimmers are linked by harmonic springs of spring constant $k$ and finite equilibrium length $b>0$. For simplicity, the springs are attached to the swimmer centers. 
In the equilibrium ground state, the colloidal molecule forms a linear straight object with the dipoles aligned along its axis. The Hamiltonian reads
\begin{align}
\mathcal{H}=&\;\frac{\mu_0m^2}{4\pi}\sum_{\substack{i,j=1\\j<i}}^3\frac{\mathbf{\hat{m}}_i\cdot\mathbf{\hat{m}}_j-3\left(\mathbf{\hat{m}}_i\cdot\mathbf{\hat{r}}_{ij}\right)\left(\mathbf{\hat{m}}_j\cdot\mathbf{\hat{r}}_{ij}\right)}{|\mathbf{r}_{ij}|^3}\nonumber\\
&+\frac{k}{2}\sum^{2}_{i=1}\left(|\mathbf{r}_{i,i+1}|-b\right)^2\label{eq:Hamilton}\,.
\end{align}
Here, $\mu_0$ denotes the vacuum permeability and $\mathbf{\hat{m}}_i=\mathbf{m}_i/m$. 
We refer to the swimmer positions as $\mathbf{R}_i=(x_i,y_i,z_i)$ in Cartesian coordinates. Then the distance vectors are given by $\mathbf{r}_{ij}=\mathbf{R}_j-\mathbf{R}_i$, where $\mathbf{\hat{r}}_{ij}=\mathbf{r}_{ij}/|\mathbf{r}_{ij}|$.

Below, we consider low-Reynolds-number dynamics in an incompressible liquid environment. Thus, hydrodynamic couplings between the swimmer bodies become important. Their resulting velocities $\mathbf{v}_i$ and angular velocities $\bm{\omega}_i$ are coupled to all forces $\mathbf{F}_j={}-\partial{\mathcal{H}}/\partial\mathbf{R}_j$ and torques $\mathbf{T}_j={}-\mathbf{\hat{m}}_j\times({\partial\mathcal{H}}/\partial\mathbf{\hat{m}}_j)$ acting on all swimmer bodies via the so-called mobility matrices \cite{Stark-Reichert}: 
\begin{equation}
\left(\begin{array}{c}\mathbf{v}_i\\ \bm{\omega}_i\end{array}\right)
=
\sum_{j=1}^3
\left(\begin{array}{cc}\bm{\mu}_{ij}^{tt}&\bm{\mu}_{ij}^{tr}\\
\bm{\mu}_{ij}^{rt}&\bm{\mu}_{ij}^{rr}\end{array}\right) 
\cdot
\left(\begin{array}{c}\mathbf{F}_j\\ \mathbf{T}_j\end{array}\right).
\end{equation}
Up to second order in $1/|\mathbf{r}_{ij}|$, the mobility matrices read 
\begin{equation}
\bm{\mu}_{ij}^{tt}=\frac{1}{8\pi\eta |\mathbf{r}_{ij}|}\left(\mathbf{I}+\mathbf{\hat{r}}_{ij}\mathbf{\hat{r}}_{ij}\right)\;\text{ for $i\ne j$}\,,\quad \bm{\mu}_{ii}^{tt}=\frac{\mathbf{I}}{6\pi\eta a},
\label{eq:Oseen}
\end{equation}
\begin{equation}
\bm{\mu}^{tr}_{ij}=\bm{\mu}^{rt}_{ij}=\frac{1}{8\pi\eta}\frac{1}{|\mathbf{r}_{ij}|^2}\mathbf{\hat{r}}_{ij}\!\!\times\;
\text{ for $i\ne j$}\,,\quad \bm{\mu}^{tr}_{ii}=\bm{\mu}^{rt}_{ii}=0,
\end{equation}
\begin{equation}
\bm{\mu}^{rr}_{ij}=0\;\;
\text{ for $i\ne j$}\,,\quad \bm{\mu}^{rr}_{ii}=\frac{\mathbf{I}}{8\pi\eta a^3},
\end{equation}
where $\eta$ is the viscosity of the surrounding fluid, $a$ is the hydrodynamic radius of the swimmer bodies, $\mathbf{I}$ denotes the identity matrix, and $\mathbf{\hat{r}}_{ij}\mathbf{\hat{r}}_{ij}$ is a dyadic product. There is no summation over $i$ and $j$ in these expressions. 

In our minimum model, the swimmers self-propel by setting the surrounding fluid into motion. 
They are driven by the drag that they experience within the self-induced fluid flow. Each swimmer acts on the fluid via a force dipole $\bm{\sigma}_i$, i.e.\ two antiparallel forces of equal magnitude, the points of action of which are slightly separated. 
We parameterize $\bm{\sigma}_i=\sigma_0\left(\mathbf{\hat{m}}_i\mathbf{\hat{m}}_i-\mathbf{I}/3\right)$ \cite{Saintillan}, thus the forces point along and rotate together with $\mathbf{\hat{m}}_i$. $\sigma_0$ sets the propulsion strength and the character of the propulsion mechanism. For $\sigma_0 <0$ we use the term ``puller'' and for $\sigma_0 >0$ the term ``pusher''. To achieve self-propulsion, the force dipoles $\bm{\sigma}_i$ are shifted out of the swimmer centers along $\mathbf{\hat{m}}_i$ by $\varepsilon> 0$ for pullers and $\varepsilon< 0$ for pushers, respectively, see Fig.~\ref{fig:sketch}. In this way, isolated swimmers always propel into the direction $\mathbf{\hat{m}}_i$. 
As a result, to the above order, one obtains ``active'' contributions to the swimmer velocities 
\cite{Pozrikidis,Jayaraman2012}
\begin{equation}
\mathbf{v}_i^a=\sum_{j=1}^3\,\bm{\mu}^{tt,a}(\mathbf{R}_i-\mathbf{R}_j-\varepsilon\mathbf{\hat{m}}_j)\,\colon\bm{\sigma}_j,
\label{eq:swimmer-vel}
\end{equation}
where $\bm{\mu}^{tt,a}(\mathbf{r})$ is a third-rank tensor of the form
\begin{equation}
\bm{\mu}^{tt,a}(\mathbf{r})=\left(-\mathbf{\hat{r}}\mathbf{I}+3\mathbf{\hat{r}}\mathbf{\hat{r}}\mathbf{\hat{r}}\right)/(8\pi\eta r^2)\,.
\label{eq:tensorD}
\end{equation}

From now on we measure all lengths in units of $b$, time $t$ in units of $6\pi \eta a/k$, and energies in units of $kb^2$. Moreover, we introduce the dimensionless parameters $\ta=a/b$, $\teps=\varepsilon/b$, $\ts=3\sigma_0\ta/2kb^2$, and $\tilde{m}^2=m^2\mu_0/{4\pi k b^5}$. Altogether, since $\mathrm{d}\mathbf{R}_i/\mathrm{d}t=\mathbf{v}_i$ and $\mathrm{d}\mathbf{\hat{m}}_i/\mathrm{d}t=\bm{\omega}_i\times\mathbf{\hat{m}}_i$, we obtain the rescaled equations of motion
\begin{align}
\frac{\mathrm{d} \mathbf{R}_i}{\mathrm{d}t} =&{}-\frac{\partial\mathcal{H}}{\partial \mathbf{R}_i} - \frac{3\ta}{4}\sum_{\substack{j=1\\j\ne i}}^3\frac{1}{|\mathbf{r}_{ij}|}\left( \mathbf{I}+\mathbf{\hat{r}}_{ij}\mathbf{\hat{r}}_{ij} \right)\cdot \frac{\partial\mathcal{H}}{\partial \mathbf{R}_j} 
+\mathbf{v}_i^a\nonumber\\ 
&{}-\frac{3\ta}{4}\sum_{\substack{j=1\\j\ne i}}^3\frac{\mathbf{\hat{r}}_{ij}}{|\mathbf{r}_{ij}|^2}\times\left(\mathbf{\hat{m}}_j\times\frac{\partial\mathcal{H}}{\partial\mathbf{\hat{m}}_j}\right)
\label{eq:eom-3d}
\end{align}
for the positions and
\begin{align}
\frac{\mathrm{d}\mathbf{\hat{m}}_i}{\mathrm{d} t} =&{}-\frac{3}{4\ta^2}\left(\mathbf{\hat{m}}_i\times\frac{\partial \mathcal{H}}{\partial \mathbf{\hat{m}}_i}\right)\times\mathbf{\hat{m}}_i\nonumber\\
&{}-\sum_{j\ne i}\frac{3 \ta}{4}\;\frac{1}{|\mathbf{r}_{ij}|^2}\left(\mathbf{\hat{r}}_{ij}\times\frac{\partial\mathcal{H}}{\partial\mathbf{R}_j}\right)\times\mathbf{\hat{m}}_i\label{eq:eom-orient}
\end{align}
for the orientations.

At rest, i.e.\ for $\ts=0$, the molecule is straight and aligned due to the magnetic 
interactions, 
see Fig.~\ref{fig:x-modes}a. When $\ts$ is switched to non-zero values, the whole molecule starts to self-propel for $\teps\neq0$. 
Artificially, we can for any value of $\ts$ delimit the molecule to the straight configuration by confining all swimmer positions and orientations to a common axis, here the $x$ axis. The distances between the swimmers and the overall speed in the resulting steadily propelling straight state are calculated numerically, 
using a fourth-order Runge-Kutta scheme. 
We set the magnetic interactions $\tm$ small enough such that a magnetic collapse due to the attractive dipole interactions does not occur.

One isolated swimmer would propel with a speed $\ts/\teps^2$. Due to the hydrodynamic interactions, the collective speed of the overall straight molecule deviates from this value. 
Moreover, for pushers ($\teps<0$, $\ts>0$) the molecule elongates, while for pullers ($\teps>0$, $\ts<0$) it contracts when compared to the non-propelling state. 
In both situations, due to the $\teps$ shift, the distance between the front and center swimmers is larger than between the center and rear swimmers.
For simplicity, we set $|\teps|=\ta$ from now on.

Our scope is to determine the stability of the resulting steady straight configuration against small perturbations, as they may arise, e.g., from imperfections in the system, thermal fluctuations, or perturbations from outside. 
%
We parameterize the swimmer positions as $\mathbf{R}_i(t)=\left(x_i(t)+\delta x_i(t), \delta y_i(t), \delta z_i(t)\right)$ and the orientations of the magnetic moments as $\mathbf{\hat{m}}_i(t)=\left(1, \dmi_y(t), \dmi_z(t)\right)$ to linear order in the deviations $ \delta x_i(t)$, $\delta y_i(t)$, $\delta z_i(t)$, $\dmi_y(t)$, and $\dmi_z(t)$ from the straight aligned configuration. Next, we linearize the system of Eqs.~(\ref{eq:eom-3d}) and (\ref{eq:eom-orient}) in these deviations. Summarizing all deviations ($i=1,2,3$) in a 15-dimensional vector $\bm{\delta}\mathbf{Q}(t)$, the resulting system of dynamic equations for the deviations has the form 
\begin{equation}
\frac{\mathrm{d}\bm{\delta}\mathbf{Q}(t)}{\mathrm{d}t}=\mathbf{M}\cdot\bm{\delta}\mathbf{Q}(t).\label{eq:eom-3d-perturbations}
\end{equation}
The coefficient matrix $\mathbf{M}$ depends on $\ts$ as it contains the swimmer separation distances in the unperturbed steady straight configuration. 
Inserting the ansatz $\bm{\delta}\mathbf{Q}(t)=\bm{\delta}\mathbf{Q}_0\exp(\lambda t)$ into Eq.~(\ref{eq:eom-3d-perturbations}), we obtain 
\begin{equation}
\mathbf{M}\cdot\bm{\delta}\mathbf{Q}_0(t)=\lambda\,\bm{\delta}\mathbf{Q}_0(t). 
\label{eq:matrix-form}
\end{equation}
Thus, the eigenvalues of $\mathbf{M}$ identify the relaxation rates of deviations from the straight configuration. The eigenvectors determine the possible corresponding deformational modes. Both were determined numerically 
\cite{eigen}.

\section{Longitudinal perturbations}
\label{sec:results}

\begin{figure}
\begin{center}
\includegraphics[width=0.8\columnwidth]{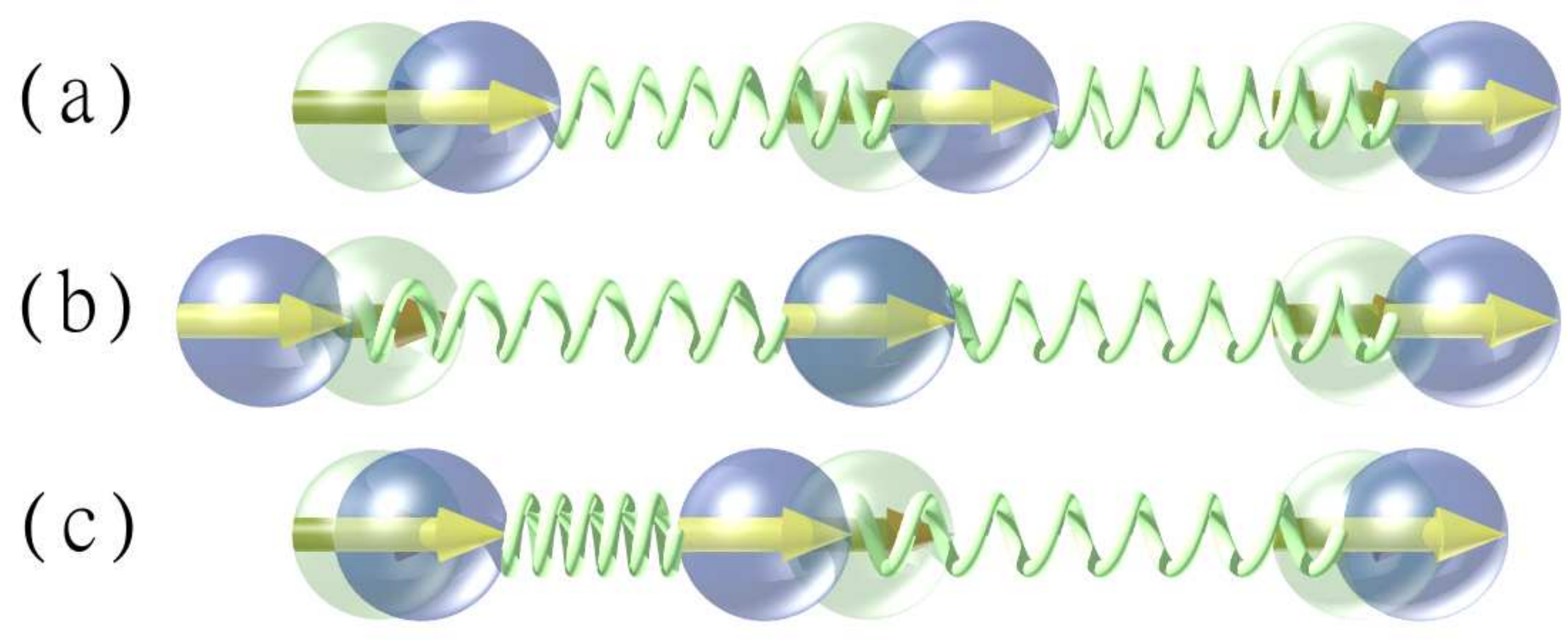}
\caption{Sketch of the one-dimensional longitudinal deformation modes in the aligned straight configuration: (a) rigid translation, (b) extension or compression, 
(c) displacement of the center swimmer with respect to the outer ones. Lighter colors 
indicate the unperturbed state for reference.
\label{fig:x-modes}}
\end{center}
\end{figure}

\begin{figure}
\includegraphics[width=\columnwidth]{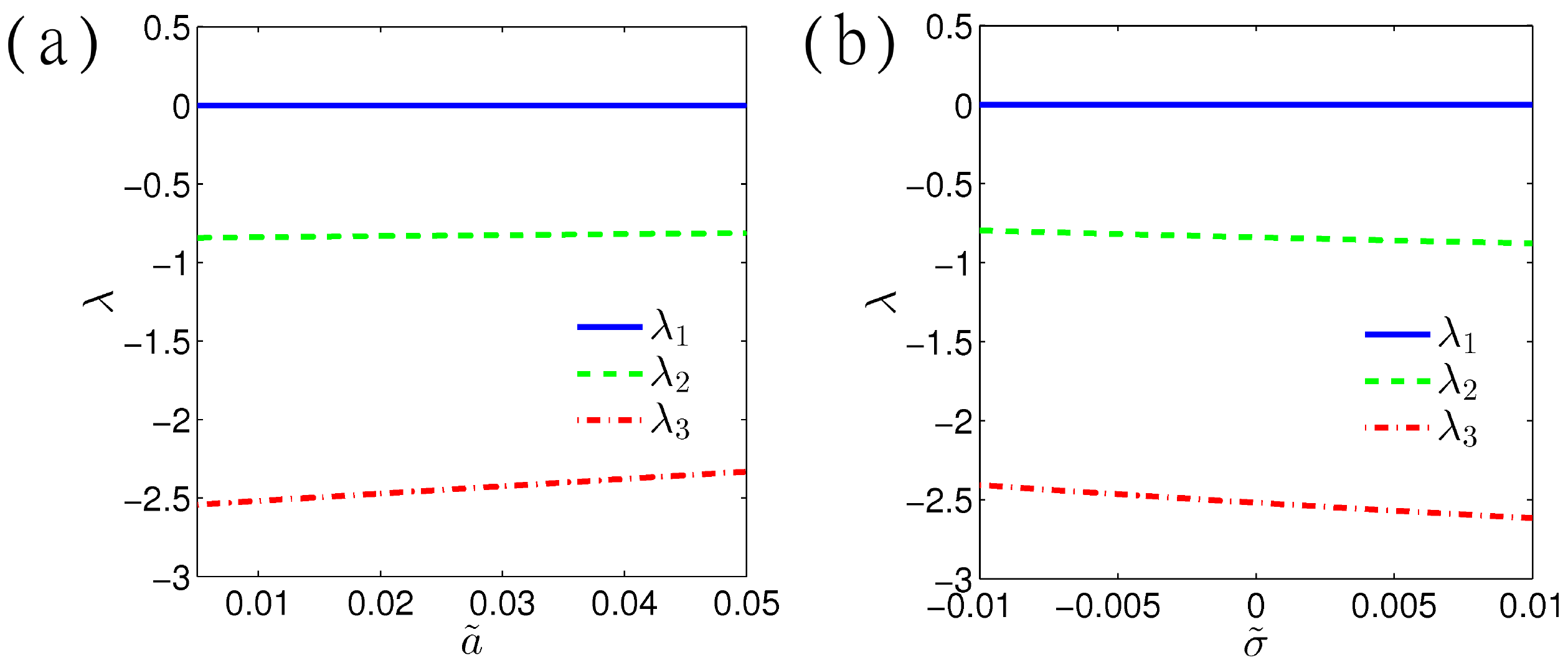}
\caption{
Influence of (a) swimmer size $\ta$ and (b) propulsion strength $\ts$ on the relaxation of longitudinally perturbed aligned straight swimmer molecules. 
The rates $\lambda_1$, $\lambda_2$, $\lambda_3$ correspond to the modes depicted in Fig.~\ref{fig:x-modes}a,b,c, respectively. Rigid translation is a zero mode, $\lambda_1=0$. Relaxation slows down with swimmer size $\ta$ and, for pullers, with propulsion strength $\ts<0$. It speeds up with 
$\ts>0$ for pushers.
[Parameters: $\tilde{m}^2=0.005$, (a) $\ts=0$, (b) $\tilde{a}=0.01$.]
}
\label{fig:1d-gamma-sigma-effect}
\end{figure}

First, we stick to the aligned straight configuration of the swimmer molecule and only consider longitudinal perturbations. That is, we set $\delta 
y_i, \delta z_i, {\dmi}_y, {\dmi}_z=0$ and only allow deviations $\delta x_i(t)$ ($i=1,2,3$). This offers a first insight into the role of the surrounding fluid and the active drive. The three resulting deformational modes (for not too high $|\ts|$) qualitatively agree with those in the absence of self-propulsion and hydrodynamic interactions, see Fig.~\ref{fig:x-modes}. 
As expected, the translational mode (Fig.~\ref{fig:x-modes}a) represents a zero mode ($\lambda_1=0$), independently of the radius $\ta$ of the swimmer bodies and the propulsion strength $\ts$, see Fig.~\ref{fig:1d-gamma-sigma-effect}. 

The relaxation of the two remaining modes, i.e.\ extension/compression of the whole swimmer molecule (Fig.~\ref{fig:x-modes}b) and displacements of the central swimmer with respect to the outer ones (Fig.~\ref{fig:x-modes}c), slows down with increasing swimmer size $\ta$. Thus, the relaxation rates $|\lambda_2|$ and $|\lambda_3|$ decrease, see Fig.~\ref{fig:1d-gamma-sigma-effect}a. Naturally, larger $\ta$ increase the friction with the surrounding fluid and enhance the hydrodynamic interaction with the other swimmers. Here, this has a decelerating effect: neighboring swimmers that tend to relax the deformation of their linking spring must move into opposite directions; then the fluid flow induced by one of the two swimmers opposes the motion of the other.


For increasing propulsion strength $|\ts|$, the relaxation of perturbations speeds up for pushers ($\ts>0$) and slows down for pullers ($\ts<0$), see Fig.~\ref{fig:1d-gamma-sigma-effect}b. 
To understand this difference, we should recall Eq.~(\ref{eq:tensorD}): the flow fields induced by the active force dipoles decay as $r^{-2}$ with distance $r$. Thus relaxations of compressed springs, involving shortened distances, have more weight in determining the dependence on $\ts$ than relaxations of extended springs. Naturally, the flow fields induced by pushers ($\ts>0$) support the separation process of two neighboring swimmers after they have come too close. This enhances the  relaxation of the linking spring. In contrast to that, the flow field induced by pullers ($\ts<0$) tends to drag the swimmers towards each other and thus hinders decompression of the spring. Then relaxation is slowed down. 
%

\section{Destabilization of the aligned straight state}

\begin{figure}
\includegraphics[width=\columnwidth]{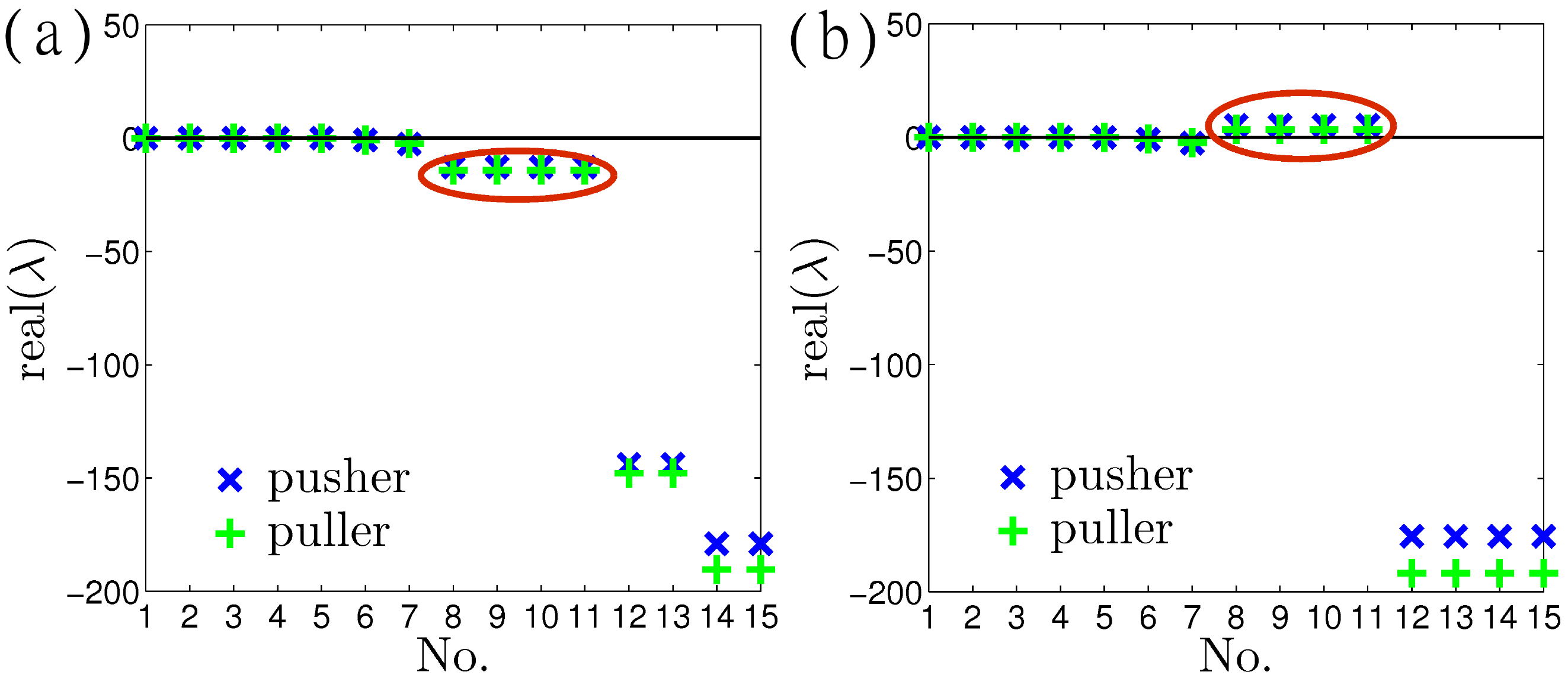}
\caption{
Real parts of the relaxation rates $\lambda$ for (a) a stable case ($|\ts|= 0.005$) and (b) an unstable situation ($|\ts|= 0.01$). In both panels, we distinguish between pushers ($\ts>0$) and pullers ($\ts<0$). We first find five zero modes corresponding to rigid translations or rotations. The next two modes are identified as the one-dimensional longitudinal modes displayed in Fig.~\ref{fig:x-modes}b,c. Both, rotations and translations into the transverse directions, appear in the remaining eight modes. Out of these, the last four are dominated by rotational perturbations and relax much faster than the translational components in the intermediate modes. Those modes leading to destabilization at high $|\ts|$ are marked by a loop. [Parameters: $\tilde{m}^2=0.005$, $\tilde{a}=0.01$.]
}
\label{fig:spectrum}
\end{figure}

Finally, we analyze the stability of the straight configuration of the swimmer molecule against all possible perturbational degrees of freedom $\bm{\delta Q}$. This leads to 15 perturbational modes, resulting from the three spatial and two orientational degrees of freedom per swimmer. 
Typical relaxation spectra, i.e.\ the real parts of the relaxation rates $\lambda$, are shown in Fig.~\ref{fig:spectrum}.%
Naturally, five zero modes representing rigid translations and rotations emerge. Next, we find that longitudinal perturbations (see Fig.~\ref{fig:x-modes}b,c) always decay. Apart from that, in the investigated regime, we observe the molecule to be stable against perturbations dominated by local rotations of the individual swimmers (the last four modes in Fig.~\ref{fig:spectrum}). Yet, with increasing propulsion strength $|\ts|$, the molecule becomes linearly unstable against four intermediate perturbational modes, signaled by positive real parts of their eigenvalues $\lambda$. They are marked by the loop in Fig.~\ref{fig:spectrum}. Actually these modes form two pairs, resulting from the fact that there are two degenerate transverse directions.

\begin{figure}
\begin{center}
\includegraphics[width=0.8\columnwidth]{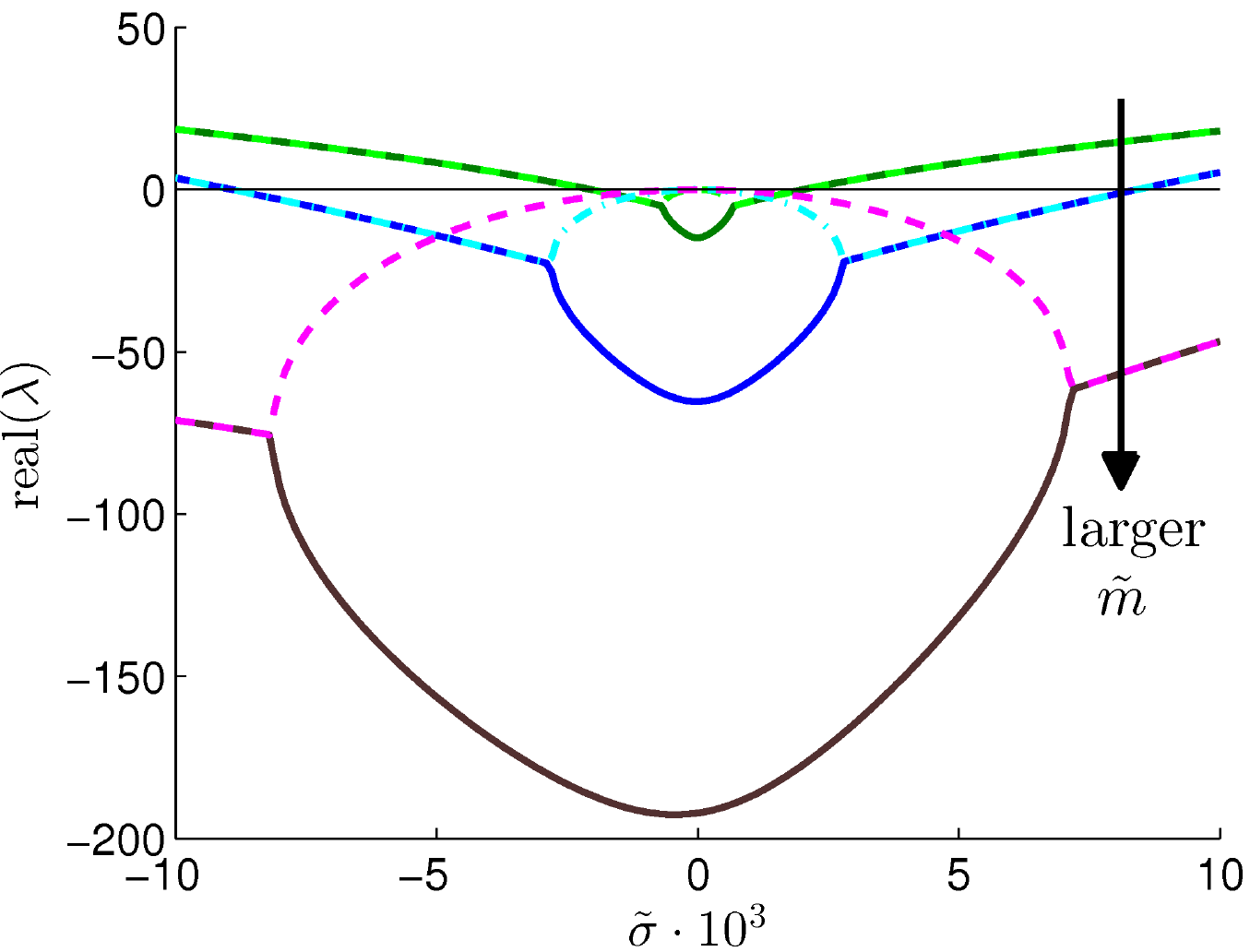}
\caption{
Real parts of the eigenvalues $\lambda$ for one pair of perturbational modes against which the straight aligned configuration first becomes unstable with increasing propulsion strength $|\ts|$. Stronger magnetic interactions, here set by $\tm^2=0.00125$, $\tm^2=0.005$, and $\tm^2=0.01125$, stabilize the straight state. Between the branching points, the relaxation rates $\lambda$ are negative and real; beyond these points, they form a pair of complex conjugate eigenvalues. [Parameters: $\ta=0.01$.]
\label{fig:hopf}}
\end{center}
\end{figure}

For one of these two degenerate pairs, we plot the real parts of the eigenvalues $\lambda$ as a function of the strength of self-propulsion $|\ts|$ in Fig.~\ref{fig:hopf}.
%
At low propulsion strength $|\ts|$, these $\lambda$ are real and negative. Thus small perturbations decay in a simple relaxation process. On the curves in Fig.~\ref{fig:hopf} this regime corresponds to the inner rounded parts. Interestingly, starting from the center of these curves, upon increase of $|\ts|$, the straight configuration is first stabilized. The upper branch, corresponding to the less stable mode, drops towards lower $\lambda<0$. At a certain $\ts$, the two branches of $\lambda<0$ meet. Beyond this point, the two relaxation rates form a complex conjugate pair. Then, the dynamic response of the swimmer molecule to the perturbations changes qualitatively. Perturbations now decay in an \textit{oscillatory} way.

Further increasing $|\ts|$ in Fig.~\ref{fig:hopf}, the real parts of the eigenvalues $\lambda$ become positive at a certain critical point. Above this threshold, the straight aligned state is linearly unstable. This happens the later the more the system is stabilized by magnetic interactions $\tilde{m}$. The instability is of an \textit{oscillatory} type.
Technically speaking, at the critical point, the real parts of the pair of complex conjugate eigenvalues change their sign. This meets the requirement for a \textit{Hopf bifurcation} scenario \cite{cross1993pattern}.

\begin{figure}
\centering
\includegraphics[width=0.85\columnwidth]{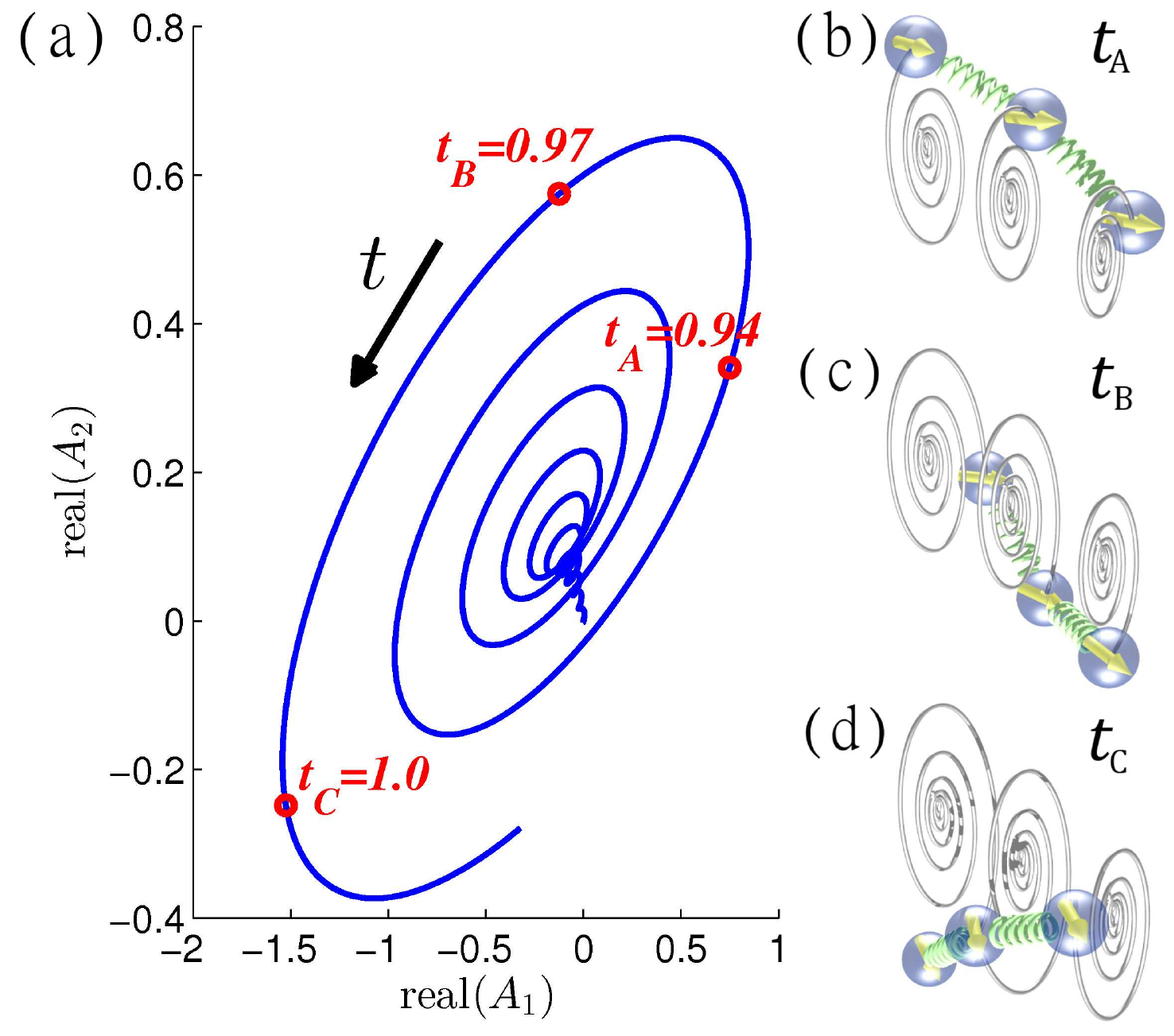}
\caption{Oscillatory linear instability of a straight initial configuration, here for the case of pushers (qualitatively the same properties are observed for pullers). (a) Tracking the real parts of the amplitudes of the two degenerate unstable pairs of modes over time, $\mathrm{real}(A_1)$ and $\mathrm{real}(A_2)$, reveals the unstable oscillatory cycle that spirals outwards. 
(b)--(d) Illustration of the destabilized configurations at the times marked in (a). Spirals indicate the real-space trajectories of the individual microswimmers during destabilization in a comoving frame. For better visualization, rotations are enlarged by a relative factor of $4$.
[Parameters: $\tilde{m}^2=0.005$, $\tilde{a}=0.01$, $\ts=0.01$.]
} 
\label{fig:pusher-amplitude-unstable}
\end{figure}

Fig.~\ref{fig:pusher-amplitude-unstable} illustrates the dynamics of the swimmer molecule during the oscillatory instability as predicted by the linear analysis. 
We numerically iterate the linearized dynamic Eq.~(\ref{eq:eom-3d-perturbations}) forward in time, starting from a weak perturbation of the straight initial state. At each time step, we project the state of the whole molecule onto the unstable complex eigenmodes. This gives their amplitudes,  
which quantifies how these modes contribute to the present configuration. In the parametric plot of Fig.~\ref{fig:pusher-amplitude-unstable}a, we track the configuration as a function of time: the abscissa and ordinate, respectively, indicate the real parts of the amplitudes of the two pairs of degenerate unstable eigenmodes. This plot demonstrates that (i) the occupation of each of the two pairs of unstable eigenmodes, given by the abscissa and ordinate, respectively, 
oscillates over time; (ii) the occupation oscillates between the two pairs of unstable eigenmodes, leading to the cycles around the origin; and (iii) the system is unstable as the cycle spirals outwards. 
Fig.~\ref{fig:pusher-amplitude-unstable}b--d illustrate snapshots of the overall configuration during one cycle. The oscillatory cycle in Fig.~\ref{fig:pusher-amplitude-unstable}a shows up as spiral-like motions of the individual swimmers.

\begin{figure}
\begin{center}
\includegraphics[width=0.8\columnwidth]{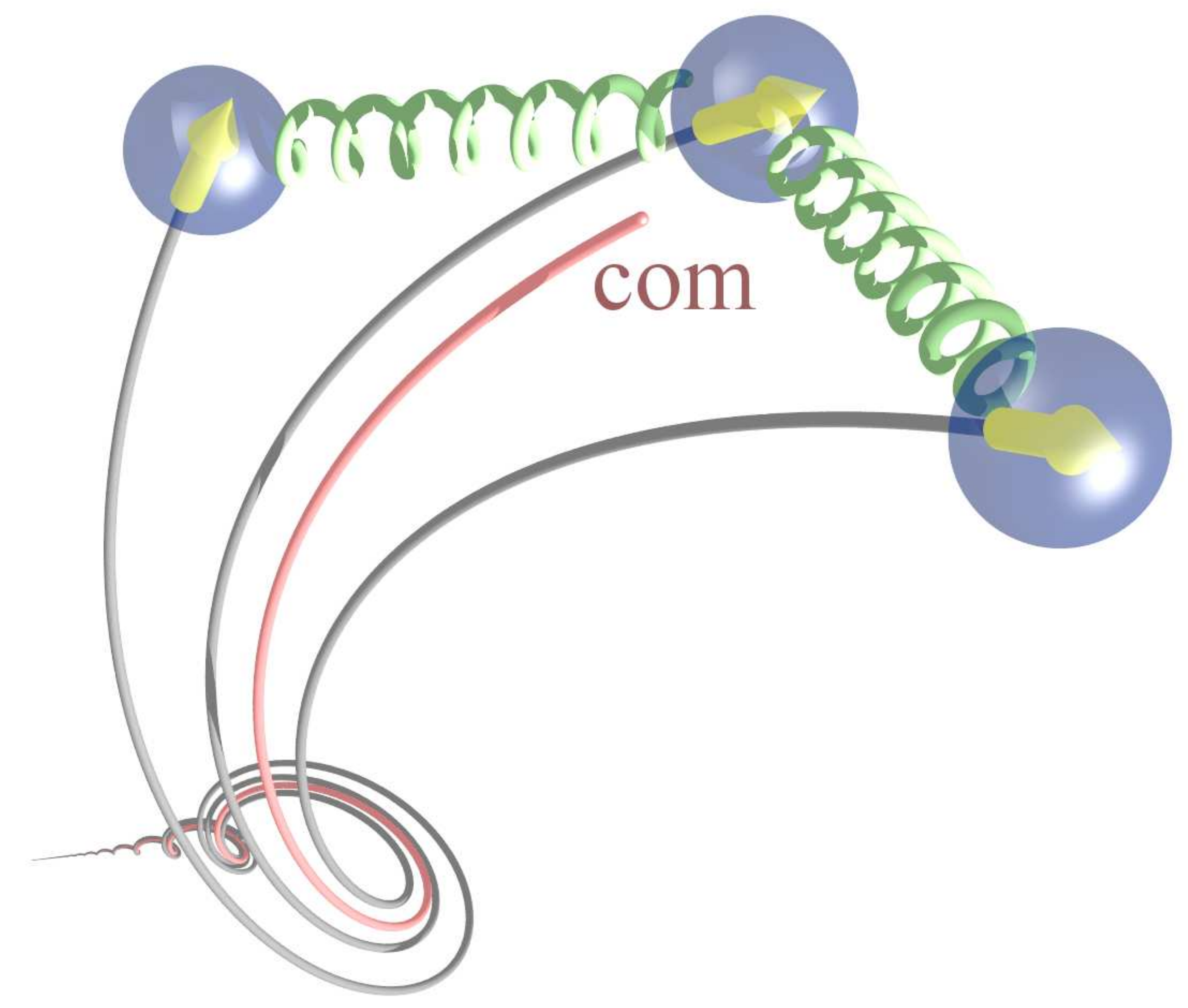}
\caption{
Example trajectory of the three-swimmer molecule in the linearly unstable regime after slight perturbation of the initially straight configuration. A corkscrew-like trajectory is observed for the individual swimmers and the center of mass (com). 
[Parameters: $\tilde{m}^2=0.005$,  $\tilde{a}=0.01$, $\ts=0.01$.] 
\label{fig:trajectory}}
\end{center}
\end{figure}

Strictly speaking, the linearized Eq.~(\ref{eq:eom-3d-perturbations}) can only predict the onset of the linear instability and describe the system behavior at low amplitudes just after destabilization. To further illustrate the motion beyond the instability, we numerically iterate the full nonlinear system of Eqs.~(\ref{eq:eom-3d}) and (\ref{eq:eom-orient}) forward in time. An example trajectory obtained in this way after slight perturbation of a straight initial configuration is depicted in Fig.~\ref{fig:trajectory}. 
There, the oscillatory instability is reflected in real space by a corkscrew-like trajectory of the whole swimmer molecule.

\begin{figure}
\begin{center}
\includegraphics[width=.85\columnwidth]{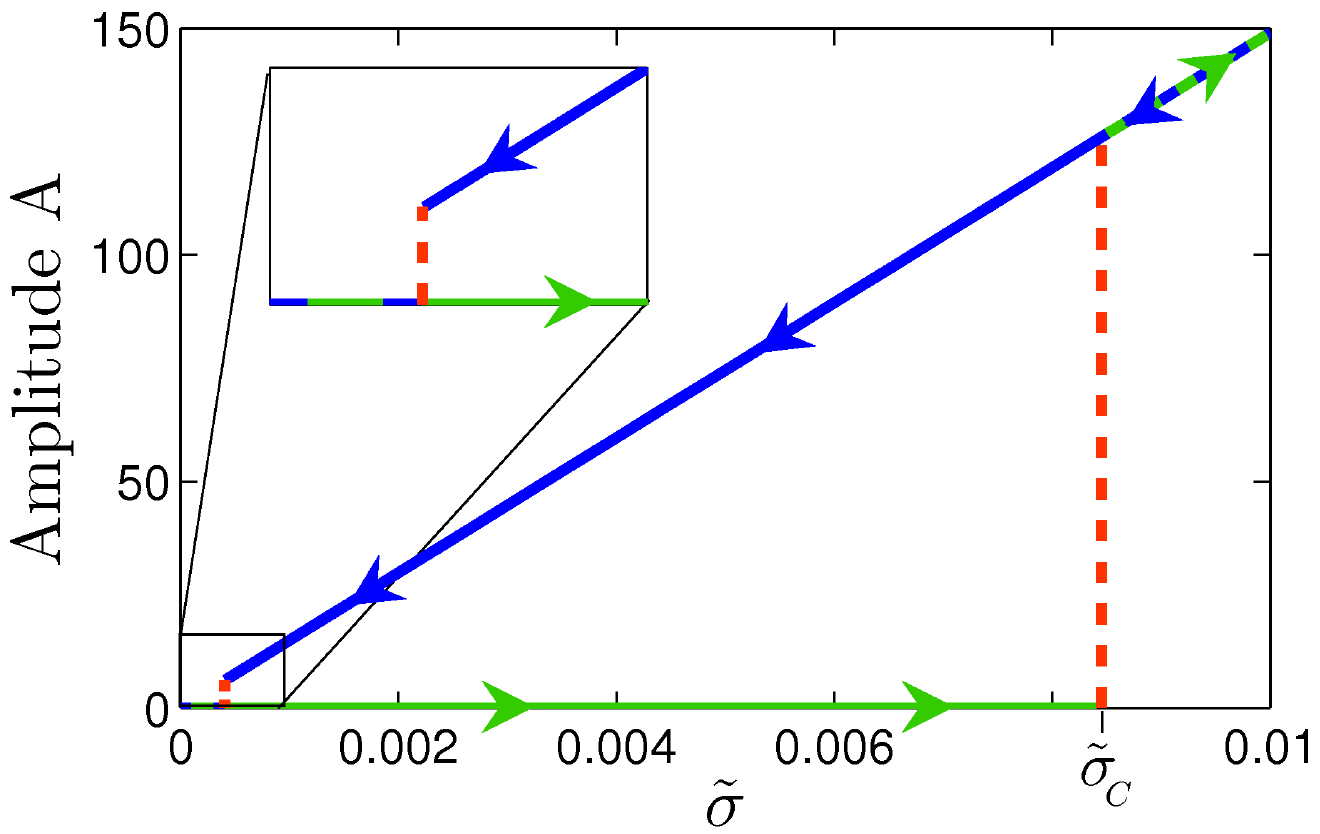}
\caption{
Hysteresis loop for the deviation from a straight configuration of the swimmer molecule, measured by the amplitude $A$ (definition see text). 
The straight configuration becomes linearly unstable at a critical point $\ts_c$, where the amplitude jumps to finite values $A\neq0$. Upon subsequent decrease of $\ts$, the amplitude jumps back to $A=0$ at $\ts<\ts_c$ as highlighted by the inset. [Parameters: $\tilde{m}^2=0.005$,  $\tilde{a}=0.01$.]
\label{fig:hysteresis}}
\end{center}
\end{figure}

Finally, we wish to clarify the nature of the bifurcation. For this purpose, we numerically iterate 
Eqs.~(\ref{eq:eom-3d}) and (\ref{eq:eom-orient}) forward in time for varying propulsion strength $\ts$. After each change in $\ts$, we wait until a steady state is reached. 
The deviation from the straight configuration is quantified by an amplitude $A=[ \sum_{i=1}^3 ( \Delta_{i,\parallel}^2+\Delta_{i,\bot}^2 ) ]^{1/2}$, where $\Delta_{i,\parallel}$ and $\Delta_{i,\bot}$ 
measure the longitudinal and transversal displacements of the swimmer bodies with respect to a steady straight configuration of the whole molecule. 
Increasing $\ts\geq0$ in Fig.~\ref{fig:hysteresis}, we observe a jump to nonzero values of $A\neq0$ at a certain critical point $\ts_c$ that agrees with the prediction of the linear stability analysis. 
Upon decreasing $\ts$ again, the swimmer molecule returns to its straight configuration only at significantly lower values $\ts<\ts_c$. 
The system shows hysteretic behavior. 
Altogether, this signals a subcritical nature of the Hopf bifurcation.

\section{Conclusions}

In this work, we have investigated the dynamic behavior of a linear magnetic microswimmer molecule. 
The colloidal molecule consists of three individual self-propelled microswimmers, connected in a linear arrangement by elastic harmonic springs. These individual swimmers hydrodynamically interact with each other, with slight variations arising from pusher or puller propulsion mechanisms. Magnetic interactions support a straight configuration of the molecule. Yet, increasing the propulsion strength, the straight configuration is destabilized. As a central result, we find that the destabilization occurs in the form of an oscillatory linear instability, in accord with a subcritical Hopf bifurcation scenario. Hysteresis is observed as a function of the propulsion strength.

Our description can be extended in many ways, for instance by addressing more than three linked self-propelled particles \cite{kaiser2015does} or different swimmer topologies, e.g.\ higher-dimensional objects \cite{kuchler2015getting} or ring-like structures \cite{messina2014self,kaiser2015active}. Individual swimmers of varying sizes and propulsion strengths, direct correlations between swimmer rotations and their mutual distances \cite{pessot2015towards}, as well as the collective behavior of many interacting molecules may be analyzed. We further hope that our predictions will stimulate experimental investigations. 
Corresponding colloidal microswimmer molecules could e.g.\ be generated by linking magnetic self-propelled Janus particles \cite{baraban2012catalytic} via DNA polymer chains \cite{dreyfus2005microscopic}. The strength of self-propulsion can be tuned in light-controlled experiments \cite{buttinoni2012active}. 
Aspects of our results may further be important for the behavior of interacting magnetotactic bacteria \cite{blakemore1975magnetotactic,frankel1979magnetite}. 
More artificially, spring-like interactions between the constituents could be mimicked by caging them in comoving optical laser traps using feedback control loops \cite{kotar2010hydrodynamic}.

\acknowledgments
The authors thank Giorgio Pessot for helpful discussions and the Deutsche Forschungsgemeinschaft for support through the SPP 1681.


%

%
%
%
%
\end{document}